\documentstyle[12pt]{article}
\newcommand{\beq}{\begin{equation}}
\newcommand{\eeq}{\end{equation}}

\newcommand{\unmezzo} {\frac{1}{2}}

\newcommand{\unquarto}{\frac{1}{4}}

\begin{document}
\pagenumbering{arabic}

\title{ON LATTICE GAS MODELS FOR DISORDERED SYSTEMS}
\author{Francesco M. RUSSO
\\[1.5em]
Dipartimento di Fisica,\\
Universit\`a di Roma {\it Tor Vergata},\\
Via della Ricerca Scientifica 1, 00133 Roma, Italy\\
and\\ INFN, Sezione di Roma}
\date{October 10, 1997}

\maketitle

\begin{abstract}

We consider a Lattice Gas model in which the sites interact via infinite-ranged
random couplings independently distributed with a Gaussian probability density.
This is the Lattice Gas analogue of the well known Sherrington-Kirkpatrick
Ising Spin Glass. We present results of replica approach in the Replica
Symmetric approximation. Even with zero-mean of the couplings a line of first
order liquid-gas transitions occurs. Replica Symmetry Breaking should give up
to a glassy transition inside the liquid phase.

\end{abstract}

\vfill
{\bf \hfill ROM2F/97/42}\\
{\bf \hfill Submitted to  Physics Letters }
\newpage

 ~ It is well known\cite{huang} that Ising Model is equivalent to Lattice Gas,
a system defined in terms of occupation variables $\tau$ taking values $0$ and
$1$. The Lattice Gas effective Hamiltonian is formally identical to Ising
one\cite{huang}. A simple change of variables ($\sigma=2\tau-1$) maps each of
the two Hamiltonians into the other, provided that Ising external field is
related to lattice gas chemical potential by $h-J=\unmezzo\mu$, where $J$ is
Ising spin-coupling related to lattice gas site-coupling $\Phi$ by
$J=\unquarto\Phi$. This reflects into a simple relation between Ising free
energy density and Lattice Gas pressure: $p=h-\unmezzo J-f$. The two systems 
have therefore the same phase diagram and the same critical behaviour (real
gases and Ising magnets are in the same universality class).

 For random systems\cite{MPV} this whole argument breaks down because the
relation between chemical potential and magnetic field involves the quenched
couplings. As we shall see in the following this reflects in new and unexpected
features for the phase diagram of the system.

  For Neural Networks the inequivalence between spin($\pm 1$) and 
occupation ($0,1$) variables was already been pointed out and analyzed,
see for example \cite{tsofei} and references therein.

Recently much effort has been devoted to develop a description of the
structural glass transition\cite{BOUMEZ,MAPARI1,MAPARI2} within the framework
of disordered systems. All these models were however based on Ising Spin
variables instead of Lattice Gas ones which would be more appropriate for
a condensed matter system. For disordered systems the two kind of variables
are not equivalent. To have a comparison term it would be useful to analyze
the properties of a mean field disordered lattice gas model.

  ~ We consider a system of $N$ sites, an occupation variable $\tau_k$ is defined
in each site $k$, $\tau_k$ can take the values $0$ or $1$. The Hamiltonian of
the system is taken to be formally identical to the SK's one\cite{SK1,SK2}. The
interaction energy between two different ($k$ and $l$) occupied sites is taken
to be $\phi_{kl}$ and the system is coupled to some external source $g$. The
total effective Hamiltonian is therefore:
\beq
 H_\phi[\tau]=-g\sum_{k=1}^{N}\tau_k-\unmezzo\sum_{l\neq k}\phi_{kl}\tau_k\tau_l
   ~.
\label{Ham}
\eeq
In magnetic language $g$ would be the external field, while for a Lattice Gas
$g$ is the sum of the chemical potential, the kinetic contribution, and
eventually an external force term.
The infinite-ranged interaction energies $\{\phi_{kl}\}$ are taken to be
quenched independent Gaussian variables with zero mean and variance $\Phi/N$.
In the following we shall take $\Phi$ as our unit of energy and set
$\Phi\equiv 1$. We also set the Boltzmann constant equal to $1$, as
a consequence $H,~f,~T,~g$ and $\phi$ are all dimensionless.
Each $\phi_{kl}$ is taken to be equally distributed and therefore each site
interacts with each other.

For a given realization of the $\phi$'s the partition function is:
\beq
 Z_\phi(\beta;g)=\sum_{\{\tau\}}e^{-\beta H\phi[\tau]}~.
\label{Zphi}
\eeq
We are interested, as usual when dealing with quenched disorder,
to evaluate the averaged free energy density:
\beq
 f(\beta;g)=-\frac{1}{\beta N}\int P[\phi]\ln Z_\phi d\phi ~,
\label{f-med}
\eeq
this will be done in the following using the replica approach\cite{MPV,EA,SK1}.

We have to calculate the averaged $n$-th power of the partition function:
\beq
 Z_n=\overline{(Z_\phi)^n}=\int(Z_\phi)^nP[\phi]d\phi ~.
\label{Zn-def}
\eeq
 For integer $n$ we get, after performing Gaussian integration:
\[
 Z_n=\sum_{\{\tau\}}\exp{\left[\sum_k\beta g\sum_a\tau_k^a+
 \frac{\beta^2}{2N}\sum_{k<l}\left(\sum_a\tau_l^a\tau_k^a\right)^2\right]}~,
\]
we can reorder the exponent and obtain:
\beq
 Z_n=\sum_{\{\tau\}}\exp\left[\beta\sum_k\left(g\sum_a\tau_k^a
 -\frac{\beta}{4N}\sum_{a,b}\tau_k^a\tau_k^b\right)\right]
 \prod_{a,b}\exp\left[\frac{4}{N\beta^2}
 \left(\frac{\beta^2}{4}\sum_k\tau_k^a\tau_k^b\right)^2\right]~.
\label{zeta-n1}
\eeq
Using Gaussian identities we rewrite (\ref{zeta-n1}) as:
\[
 Z_n=\sum_{\{\tau\}}\exp\left[\beta\sum_k\left(g\sum_a\tau_k^a
 -\frac{\beta}{4N}\sum_{a,b}\tau_k^a\tau_k^b\right)\right]
 \times
\]
\beq
 \times
 \prod_{a,b}\int\left(\frac{N\beta^2}{4\pi}\right)^\unmezzo
 \exp\left(-\frac{N}{4}\beta^2 Q_{ab}^2+
 \unmezzo\beta^2\sum_k Q_{ab}\tau_k^a\tau_k^b\right)dQ_{ab}~,
\label{zeta-n2}
\eeq
reordering the exponentials and defining:
\beq
 H_Q[\tau]=-\unmezzo\beta^2\sum_{a,b}
 \left(Q_{a,b}-\frac{1}{2N}\right)\tau^a\tau^b
 -\beta g\sum_a\tau^a ~,
\label{HQ[tau]}
\eeq
\beq
 A[Q]=\frac{\beta^2}{4}\sum_{a,b}Q_{ab}^2 ~
 -\ln\left[\sum_{\{\tau\}}e^{-H_Q[\tau]}\right]~,
\label{A[Q]}
\eeq
we finally get:
\beq
 Z_n(\beta;g)=\left(\frac{N\beta^2}{4\pi}\right)^\frac{n^2}{2}
 \int e^{-NA[Q]}~d^{n^2}Q ~.
\label{zeta-n3}
\eeq
 The averaged free energy density is given by:
\beq
 f(\beta;g)=\lim_{N\to\infty}\lim_{n\to0}
 -\frac{1}{\beta nN}\ln Z_n(\beta;g)~.
\label{free-lim}
\eeq
In the thermodynamic limit the integral can be estimated by maximizing
the integrand, this yields:
\beq
 f_n(\beta;g)=\lim_{N\to\infty}-\frac{1}{\beta nN}\ln Z_n(\beta;g)
 =\frac{1}{\beta n}\inf_Q\{A[Q]\}~.
\label{effe-n}
\eeq
The extremum is determined from the saddle point equation:
\beq
 \frac{\partial A}{\partial Q_{ab}}=
 \frac{\beta^2}{2}Q_{ab}-\frac{\beta^2}{2}
 \frac{\sum_\tau\tau^a\tau^b e^{-H_Q[\tau]}}
 {\sum_\tau e^{-H_Q[\tau]}}=0 ~,
\label{Qsp}
\eeq
that may be rewritten as $Q_{ab}=\langle\tau^a\tau^b\rangle_Q$.

 As a first stage we consider saddle points that are symmetric under
the Replica Group\cite{MPV}. Setting $Q_{ab}=q+b\delta_{ab}$ we can write
$\sum_{ab}Q_{ab}\tau^a\tau^b=q\left(\sum_a\tau^a\right)^2+b\sum_a\tau_a$ and
$\sum_{ab}Q_{ab}^2=n(q+b)^2+n(n-1)q^2$, substitution in (\ref{A[Q]}) and
extraction of the $n\to0$ limit then yields:
\beq
 f=\unquarto\beta b(2q+b)-(2\pi\beta^2)^{-\unmezzo}\int_{-\infty}^{\infty}
 \ln\left[1+e^{\beta(\alpha+z\sqrt{q})}\right]e^{-\unmezzo z^2}~dz~.
\label{f-sim}
\eeq
In equation (\ref{f-sim}) we set $\alpha=g+\unmezzo\beta b$, and the matrix
elements satisfy the coupled equations:
\[
 \rho\equiv q+b=(2\pi)^{-\unmezzo}\int
 \left[1+e^{\beta(\alpha+z\sqrt{q})}\right]^{-1}e^{-\unmezzo z^2}~dz
\]
\beq
 q=(2\pi)^{-\unmezzo}\int
 \left[1+e^{\beta(\alpha+z\sqrt{q})}\right]^{-2}e^{-\unmezzo z^2}~dz ~.
\label{Qsp-sim}
\eeq
As can be seen following the line of \cite{SK1,SK2} the physical
significance of $\rho$ and $q$ is:
\beq
 \rho=\overline{\langle\tau\rangle}~~~~~~~~~~
 q=\overline{\langle\tau\rangle^2}~,
\label{sign-sim}
\eeq
where, following the notations of \cite{MPV}, a bar denotes
the average over quenched disorder.

 For $\beta=0$ we have $\rho=\unmezzo$ and $q=\unquarto$, as we expect from
their physical significance. In the high temperature regime we can solve
(\ref{Qsp-sim}) by expansion in powers of $\beta$, this yields:
\[
 \rho=\unmezzo+\unquarto\beta g+\frac{1}{32}\beta^2
\]
\[
 q=\unquarto+\unquarto\beta g+
 \frac{1}{16}\left(\frac{3}{4}+g^2\right)\beta^2 ~.
\]

We have numerically solved equations (\ref{Qsp-sim}) for several values of $T$
and $g$. We find a line of first-order phase transitions. Such line ends in a
second-order transition point, for $T\simeq0.22$ and $g\simeq-0.7$, where the
linear response function $\chi=\partial{\rho}/\partial{g}$ (the
"susceptibility") is found to be divergent.

Let us define
\[
 \gamma_0\equiv\lim_{T\to0}\beta(\rho-q)=\lim_{T\to0}\beta b  ~~,
\]
for $T\to0$ we have $\rho=q+\gamma T$ and thus $\alpha=g+\unmezzo\gamma_0$.
  ~For $\beta\gg1$ the solution of (\ref{Qsp-sim}) can behave in two different
ways depending on which side of the first order transition we consider. We call
$g_0=-0.63633$ the transition point at zero temperature. In the range $g<g_0$,~
$q_0~(\equiv q(T;g)|_{T=0})$ and $\gamma_0(g)$ are identically zero (details
will be presented elsewhere\cite{RUSSO1}), we find that all their temperature 
derivatives vanish for $T\to0$, $q$ vanishes as $e^{\beta g}$ ($g<0$) and
$\gamma~(\equiv\beta(\rho-q))$ as $\beta e^{\beta g}$. Elsewhere if $g>g_0$,
they depend linearly on $T$, indeed we get:
\[
 q=q_0-\frac{q_0\gamma_0(q_0+\unmezzo\gamma_0\alpha_0)}
 {(q_0+\unmezzo\gamma_0\alpha_0)^2+\frac{1}{4}\gamma_0(q_0-\alpha_0^2)}~T
\]
\[
 \gamma=\gamma_0+\frac{\unmezzo\gamma_0^2(q_0-\alpha_0^2)}
 {(q_0+\unmezzo\gamma_0\alpha_0)^2+\frac{1}{4}\gamma_0(q_0-\alpha_0^2)}~T
\]
   ~
\[
 \rho=q+\gamma T ~.
\]

Next we look to thermodynamical functions, the internal energy and entropy
densities are given by:
\[
 u=-(g+\gamma_0)q
\]
\beq
 s=-\frac{1}{4}\gamma_0^2 ~.
\label{entr-T0}
\eeq
The entropy (\ref{entr-T0}) is negative in the range $g>g_0$ where $\gamma_0$
is different from zero, so we should expect the Replica Symmetry to be broken
in this region. We stress that differently from SK case we have a region in
which the Replica Symmetric Solution remains physical down to zero temperature.
This can be related to some exact results on the ground state energy of the
model\cite{RUSSO1}. The maximum absolute value of the zero temperature entropy
is at $g=g_0^+$ where it takes the value $0.101$ and it strongly decrease for
higher values of $g$ (e.g. at $g=1 ~~ s=-0.011$).
  ~ This solution is clearly unphysical (for $g>g_0$) at low temperature.
This should be considered a clear signal\cite{MPV} of Replica Symmetry
Breaking (RSB).  

Now we briefly compare our results with the ones of Sherrington and Kirkpatrick.
In SK case the zero-field Hamiltonian has a {\em global $Z_2$ symmetry\/} and
all thermodynamical functions are either even or odd in the external field. SK
model can have a (field-driven) {\em first-order phase transition\/} only if a
strong enough  ferromagnetic part (i.e. a non-zero mean) is added to the random
coupling. Such a transition is indeed related to the breaking of the global
$Z_2$ symmetry induced (as in the homogeneous case) by a ferromagnetic
coupling. Moreover, at each value of external field, SK's Replica Symmetric
Solution always becomes unphysical for low enough temperature.

It is apparent how our picture is different from the usual one. We have
considered the case of purely (zero-mean) random interactions and our
Hamiltonian has no $Z_2$ symmetry. The {\em Replica Symmetric Solution\/} of
our system exhibits a line of first-order phase transition points ending with a
second-order transition. This feature is expected to be robust to {\em Replica
Symmetry Breaking\/}, as should be confirmed by the stability
analysis\cite{DEALM,RUSSO1}. Two phases co-exist along the transition line and
in one of them the Replica Symmetry should remain {\em exact down to zero
temperature.\/} Following the conventional Lattice Gas interpretation this
two phases should be called `gas' and `liquid'. The gas-like phase is that
which is replica-stable at all temperature. When Replica Symmetry Breaking
sets up the liquid-like phase can give up to a glassy state. This idea will
be put forward elsewhere\cite{RUSSO1}.

 I am very grateful to Prof. Giorgio Parisi for the help he gave in this work
with his suggestions and experience, and also for his careful and critical
reading of the original manuscript.
I acknowledge Prof. Enzo Marinari and Dr. Felix Ritort for useful  and
stimulating discussion.  I wish acknowledge INFN, "Sezione di Roma {\em Tor
Vergata\/}", for support received during this work.


\vfill
\end{document}